\documentclass[manuscript, authorversion]{acmart}
\pdfoutput=1
\usepackage{ragged2e}

\usepackage{type1cm} % type1 computer modern font
\usepackage{graphicx} % advanced figures
\usepackage{xspace} % fix space in macros
\usepackage{balance} % to better equalize the last page
\usepackage{booktabs} % nicer tables
\usepackage{multirow} % multi rows for tables
\usepackage[font={bf}, tableposition=top]{caption} % captions on top for tables
\usepackage{subcaption} % subfloats
\usepackage{bold-extra} % bold + {small capital, italic}
\usepackage[vlined,linesnumbered,ruled,noend]{algorithm2e} % algorithms
\usepackage{microtype} % compress text
\usepackage{siunitx} % \num for decimal grouping
\usepackage{xfrac} % nicer slanted fractions
\usepackage{mathtools} % amsmath++
\PassOptionsToPackage{hyphens}{url} % acmart compatibility
\PassOptionsToPackage{bookmarks, pdftex, colorlinks=true, pagebackref=true, backref=page}{hyperref} % acmart compatibility
\PassOptionsToPackage{square,numbers}{natbib} % acmart compatibility
\usepackage{cleveref} % smart references
\usepackage[hyperpageref]{backref} % back references
\usepackage{hyphenat} % name in single line
\usepackage[show]{chato-notes} % notes

\usepackage{CJKutf8} % Japanese

\newenvironment{squishlist}
{\begin{list}{$\bullet$}
 {\setlength{\itemsep}{0pt}
     \setlength{\parsep}{3pt}
     \setlength{\topsep}{3pt}
     \setlength{\partopsep}{0pt}
     \setlength{\leftmargin}{1.5em}
     \setlength{\labelwidth}{1em}
     \setlength{\labelsep}{0.5em} } }
{\end{list}}

\renewcommand*\backref[1]{\ifx#1\relax \else (Cited on #1) \fi}

\graphicspath{{images}}

\setcopyright{none}
\renewcommand\footnotetextcopyrightpermission[1]{} % removes footnote with conference information in first column
\acmConference[]{}{}{}
\begin{document}

%% The "title" command has an optional parameter,
%% allowing the author to define a "short title" to be used in page headers.
\title{What we can learn from TikTok through its Research API}

%% The "author" command and its associated commands are used to define
%% the authors and their affiliations.
%% Of note is the shared affiliation of the first two authors, and the
%% "authornote" and "authornotemark" commands
%% used to denote shared contribution to the research.

\author{Francesco Corso}
\affiliation{%
 \institution{Politecnico di Milano \& CENTAI}
 \city{Milan}
 \country{Italy}}
 \email{francesco.corso@polimi.it}

\author{Francesco Pierri}
\affiliation{%
\institution{Politecnico di Milano}
 \city{Milan}
 \country{Italy}}
 \email{francesco.pierri@polimi.it}

\author{\texorpdfstring{\nohyphens{Gianmarco~De~Francisci~Morales}}{Gianmarco De Francisci Morales}}
\affiliation{%
  \institution{CENTAI}
 \city{Turin}
 \country{Italy}}
\email{gdfm@acm.org}

\renewcommand{\shortauthors}{Corso, Pierri, \& De Francisci Morales}

%% The abstract is a short summary of the work to be presented in the
%% article.
\begin{abstract}
TikTok is a social media platform that has gained immense popularity over the last few years, particularly among younger demographics, due to the viral trends and challenges shared worldwide.
The recent release of a free Research API opens the door to collecting data on posted videos, associated comments, and user activities.
Our study focuses on evaluating the reliability of the results returned by the Research API, by collecting and analyzing a random sample of TikTok videos posted in a span of 6 years.
Our preliminary results are instrumental for future research that aims to study the platform, highlighting caveats on the geographical distribution of videos and on the global prevalence of viral and conspiratorial hashtags.
\end{abstract}

%%
%% The code below is generated by the tool at http://dl.acm.org/ccs.cfm.
%% Please copy and paste the code instead of the example below.
%%
\begin{CCSXML}
<ccs2012>
   <concept>
       <concept_id>10002951.10003317</concept_id>
       <concept_desc>Information systems~Information retrieval</concept_desc>
       <concept_significance>300</concept_significance>
       </concept>
   <concept>
       <concept_id>10002951.10003317.10003359.10011699</concept_id>
       <concept_desc>Information systems~Presentation of retrieval results</concept_desc>
       <concept_significance>300</concept_significance>
       </concept>
   <concept>
       <concept_id>10002951.10003260.10003282.10003292</concept_id>
       <concept_desc>Information systems~Social networks</concept_desc>
       <concept_significance>500</concept_significance>
       </concept>
 </ccs2012>
\end{CCSXML}

\ccsdesc[300]{Information systems~Information retrieval}
\ccsdesc[300]{Information systems~Presentation of retrieval results}
\ccsdesc[500]{Information systems~Social networks}

%% Keywords. The author(s) should pick words that accurately describe
%% the work being presented. Separate the keywords with commas.
\keywords{online social networks, API, TikTok}

%%
%% This command processes the author and affiliation and title
%% information and builds the first part of the formatted document.

\maketitle

\section{Introduction}
TikTok is a social media platform for sharing short-form videos, known for its wide range of user-generated content, including lip-syncing, comedy, talent displays, and more.
It has seen a steep increase in popularity, becoming one of the most prominent social media platforms on the Internet, with over 1 billion monthly active users and millions of videos posted every day around the world.\footnote{\url{https://www.businessofapps.com/data/tik-tok-statistics/} accessed on 26/01/2024}
% It allows users to create and share short videos often accompanied by music or other audio clips, and it provides editing features such as filters, effects, and soundtracks. 

TikTok has recently released a public Research API,\footnote{\url{https://developers.tiktok.com/products/research-api} accessed on 24/01/2024.} to which researchers can apply for access to gather data on videos, users, and comments via three respective primary endpoints.
Such data availability initiative follows the examples of other platforms, such as Facebook\footnote{\url{https://www.crowdtangle.com}} and Twitter,\footnote{\url{https://developer.twitter.com/en/use-cases/do-research}} which have been compelled to open to their data to researchers interested in studying the integrity of digital environments~\cite{bar2023auditing}, especially under the pressure from the EU Digital Services Act.\footnote{\url{https://digital-strategy.ec.europa.eu/en/policies/digital-services-act-package} accessed on 26/01/2024.}
 
In the current post-API era~\cite{freelon2018computational}, where most of the once-free APIs have been closed or converted to paid services, TikTok is still a relatively new frontier, even though the platform has been online for over five years.
So far, the main approach to obtain data from TikTok has been to scrape and collect videos manually, as done, for instance, by \citet{guinaudeau_fifteen_2022} who studied political videos in the US, and showed differences in the activity of users of both TikTok and YouTube. 
Other works, such as the one by \citet{medina_serrano_dancing_2020}, also focused on US political discussions on TikTok, but employed a more content-based approach, and used wide-scale ML models on the video or the audio track of the videos they scraped.
Similarly, other research described the usage of TikTok by organizations to communicate safety measures, best practices, and news during the pandemic~\cite{li_communicating_2021,ostrovsky2020tiktok}, and the impact of soft moderation labels employed by the platform for videos related to the COVID-19 pandemic~\cite{ling2023learn}.
\citet{pera2023shifting} compared TikTok and YouTube, this time by using the TikTok Research API to look for climate change-related videos.
\citet{klug2021trick} used a mixed-method approach to investigate the common assumptions of users about the TikTok recommendation algorithm.
Lastly, a work similar to ours is that by \citet{mcgrady_dialing_2023}, who focused on gathering a random sample from social media (YouTube) and estimated the total number of videos present on the platform.

In this paper, we collect and analyze a random sample of TikTok videos posted in the 6 years spanning from January 2018 to December 2023 by means of TikTok Research API.
We ask the following research question.
\begin{squishlist}
\item[\textbf{RQ1:}] \textbf{What view of TikTok do we get through the lens of its Research API?}
\end{squishlist}

We provide a series of quantitative analyses on the data returned by the Research API and explore the potential implications for research relying on such a tool.
By using repeated calls to the API, we build a worldwide random sample of over 500k videos (stratified per month) and analyze engagement metrics such as likes, shares, comments, and views.
We highlight the temporal growth of the platform and show that the user base is dominated by Asian countries, with the USA as the only Western country in the top 10 in terms of shared videos.
Lastly, we underline the effects of viral hashtags on driving engagement around videos that use the specific ``For you'' functionality of the platform, and offer an outlook on the prevalence of hashtags related to conspiracy theories.

\section{Methods}
\label{sec:methods}
Our main constraint and objective for this research is to use exclusively the official TikTok Research API, which has severe restrictions on usage and availability.
In particular, each API research organization has a quota of \num{1000} available requests per day.
Since each request can return up to a maximum of \num{100} items, the resulting theoretical limit of the maximum number of elements available each day is \num{100000}.
Given our aim of studying the geographical distribution of videos, we use a query that contains all the region codes described in the TikTok API documentation (note that Canada is not available by default in the API).\footnote{ \label{video_api}\url{https://developers.tiktok.com/doc/research-api-specs-query-videos}} %accessed on 24/01/2024}
We use monthly queries since the maximum width of the time frame allowed for data collection by the API is 30 days.
To meet the constraints imposed by the Terms of Service\footnote{\url{https://www.tiktok.com/legal/page/global/terms-of-service-research-api/en} accessed on 24/01/2024} we fix our collection quota to \num{1000} videos per month.
Our sample thus aims to be stratified and have a uniform number of randomly sampled videos for each month in the period of our study.
We send \num{10} requests, each of \num{100} videos, to the \texttt{/video/query} endpoint for each month from January 2018 to December 2023 (72 months).
This yields a theoretical 72k items per day of extraction by using \num{720} of the daily \num{1000} queries available.

This collection process was run for \num{15} consecutive days from January 17th 2024 to January 31st 2024.
If the maximum quota were reached each day, it would result in a dataset of over 1 million items, with 15k videos per month in the time frame of the study.
Our methodology for data collection adheres to ethical standards as we do not try to deanonymize users.
In addition, TikTok users have explicitly consented to the Terms of Service, which include the acknowledgment and approval of the transfer of personal data through the API.\footnote{\url{https://www.tiktok.com/legal/page/eea/privacy-policy/en} accessed on 24/01}
\section{Results}
\subsection{Evaluation of the API}

\begin{figure}[!t]
    \centering
    \includegraphics[width=1\linewidth]{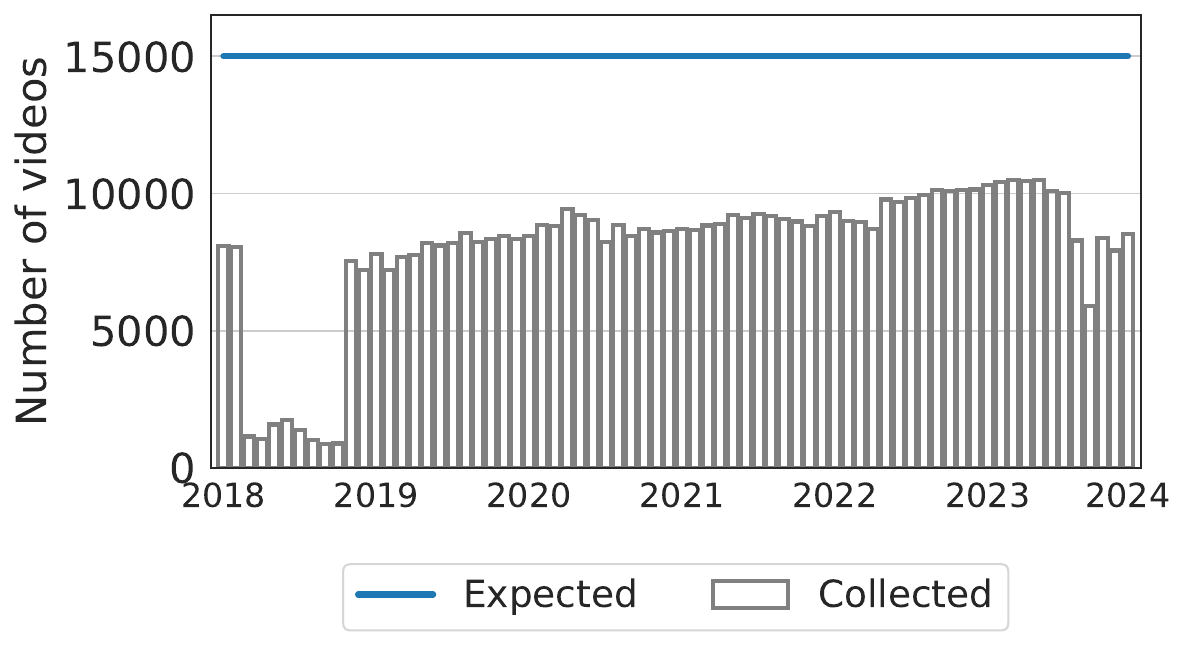}
    \vspace{-\baselineskip}
    \caption{Time series of the data collection. The blue line represents the theoretical quota (maximum number of videos obtainable with the given number of API calls), while the histogram shows the obtained quota per month.}
    \label{fig:rq1_ts}
\end{figure}

\begin{figure}
    \centering
    \includegraphics[width=1\linewidth]{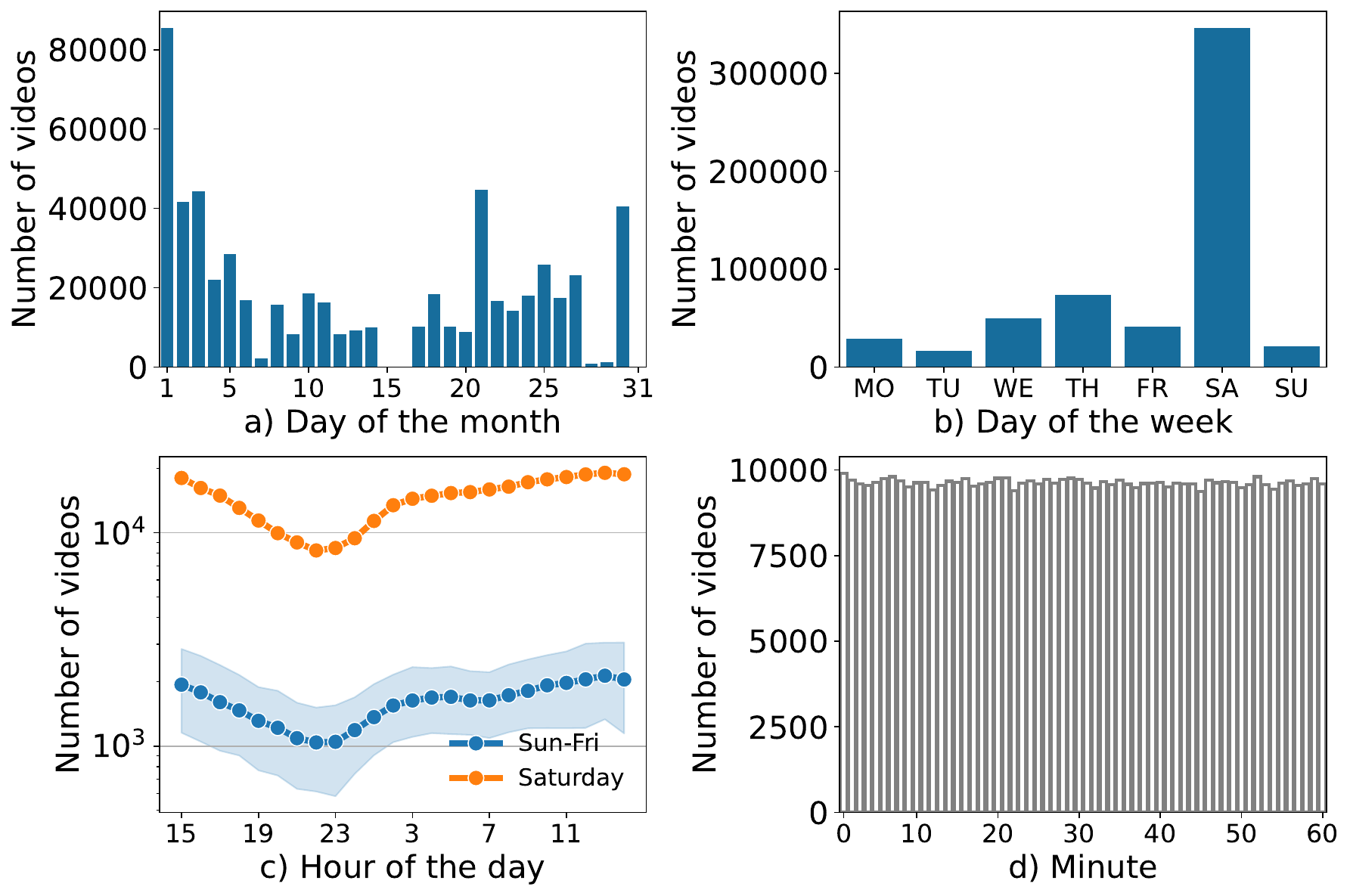}
    \caption{Number of videos posted (a) for each day of the month, (b) for each day of the week, (c) for each hour of the day (UTC), and (d) for each minute of the hour.
    The API shows a bias at the daily level, but not at the minute level.}
    \label{fig:rq4_time}
\end{figure}

\Cref{fig:rq1_ts} shows the theoretical and real number of videos obtained from the data collection process described in \Cref{sec:methods}.
The API failed to meet the required quotas, delivering at most 65\% (out of 72k) of the requested videos, a number which is in line with the data persistence of other social networks~\cite{elmas2023impact}.
At the end of the data collection process, we obtained a total of \num{577517} videos instead of the estimated $1$+ million.
The number of distinct users in the collection is almost the same as the number of videos, with only \num{0.34}\% repeated users in the sample.
This result is expected given the extremely large number of users on the platform (over 1 billion).

The API documentation\footnotemark[6] explains that the requested quotas might not be met when videos marked as private or deleted appear in the response.
These videos are not returned but are still counted by their internal system as part of the query result.
This fraction of unavailable videos can thus give us an indication of the proportion of unavailable content on TikTok, month by month.
There is a growing trend in the number of returned videos, which finds its peak in March 2023,
probably because older videos are more likely to be deleted by users.
There is also a large gap in the period from March 2018 to December 2018 where the returned data points barely surpassed 500 items per month.
A similar but smaller drop appears around July 2023.
We do not know the causes behind the missing videos in 2018, but we can speculate that it is possibly due to an error in the internal systems of the API.
For this reason, we exclude the year 2018 from the following analyses.
\subsection{Temporal Bias}

Here, we investigate temporal patterns in the videos present in our sample.
First, we analyze the frequency of appearance of all the different days of the month. In \Cref{fig:rq4_time}(a), we show the cumulative number of videos posted for each day of the month, indicating that our data is not a perfect random sample, since the distribution is not uniform across all days of the month.
For instance, we observe an unusual amount of videos posted on the first day of the month along with some missing days (e.g., the 15th).
\Cref{fig:rq4_time}(b) shows the number of videos posted for each day of the week. Saturday is the day when the majority of the videos are posted (>55\%) followed by Thursday and Wednesday. 
Both of these observed facts are probably due to a malfunctioning of the Research API internal mechanisms, as we do not have reasonable evidence showing that these two phenomena are generated by user behaviour on the platform.
\Cref{fig:rq4_time}(c) shows the number of videos posted for every hour (UTC zone).
We plot two time series to show the difference in volumes for Saturday compared to the rest of the days of the week.
We find similar behavior to what was shown on other social media platforms like Twitter \cite{pfeffer_just_2023}, where there is a peak of posted videos in the early afternoon. Despite the fact that videos posted on Saturday are three times larger than on other days of the week, the distribution pattern of the posting time remains very similar.
Finally, \Cref{fig:rq4_time}(d) shows the distribution of the minutes of the posting time.
We find a different behavior compared to what has been evidenced by \citet{pfeffer_just_2023} on Twitter (now X), where \num{15}\% of the data they collected was generated in the first minute of the hour in which they were posted.
The distribution on TikTok is instead an almost uniform across all minutes. 
\citeauthor{pfeffer_just_2023} suggested their result was due to bots' activity and programmed tweets.
TikTok also allows scheduling video releases in advance, but only to Creators and Business accounts,\footnote{\url{https://www.tiktok.com/business/en-US/blog/introducing-video-scheduler-now-you-can-plan-tiktoks-in-advance} accessed on 20/02/2024} which are a small minority of the user-base of the platform.
Thus this functionality seemingly does not influence the pattern of the posting times.

\subsection{Distribution of interactions}
\begin{figure}[!t]
    \centering
    \includegraphics[width=.99\linewidth]{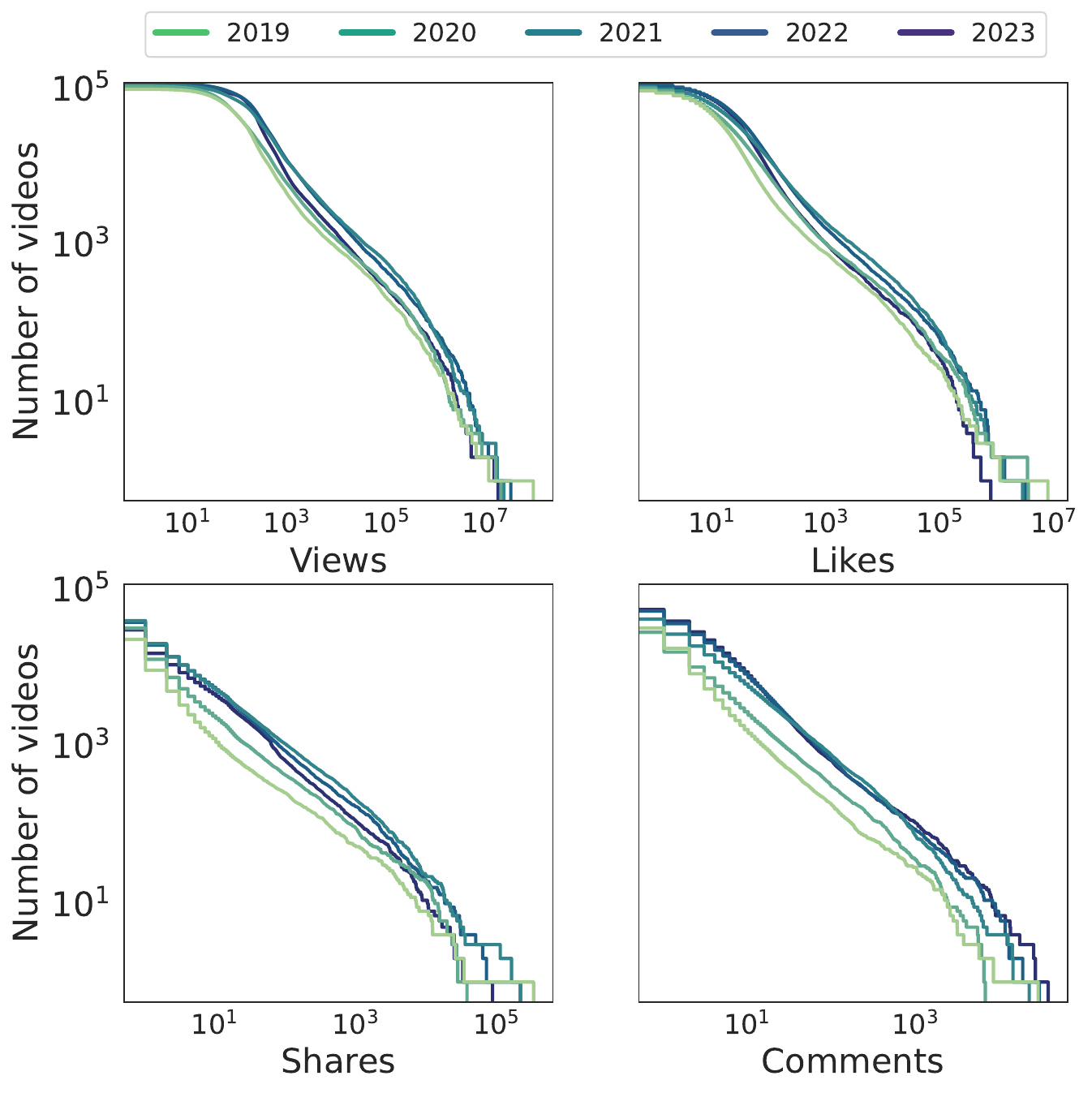}
    \vspace{-\baselineskip}
    \caption{CCDFs of the four main interactions on TikTok: number of views, of likes, of shares, and of comments for videos per year. All the features have a heavy-tailed distribution. The yearly platform growth is evident in the shift to the right of each feature. Axes are on a logarithmic scale.}
    \label{fig:rq2_ccdf}
    \vspace{-\baselineskip}
\end{figure}

Let us now focus on the interaction indicators available in our sample.
% are distributed in order to describe the qualities of the average type of content published on the platform.
\Cref{fig:rq2_ccdf} shows the complementary cumulative distribution function (CCDF) of the four features available on the API: the number of views, likes, shares, and comments for each video.
% We plot on the y-axis the number of videos in our sample that have the corresponding number of views, likes, shares, or comments on the x-axis.
The plot shows a scaling behavior typical of social networks~\cite{adamic2001search}. There has been a progressive increase in the median values for views and likes over the years, as these indicators follow the platform's growth.
The order of the subplots is increasing in `strength' of interaction~\cite{medina_serrano_dancing_2020}, with the lowest defined as the visualization of the video, the second with a like, the third with a share, and the fourth with a comment.
Indeed, the latter two have maximum values of two orders of magnitude lower than views and likes.

\subsection{Region prevalence}

\begin{figure}[!t]
    \centering
    \includegraphics[width=1\linewidth]{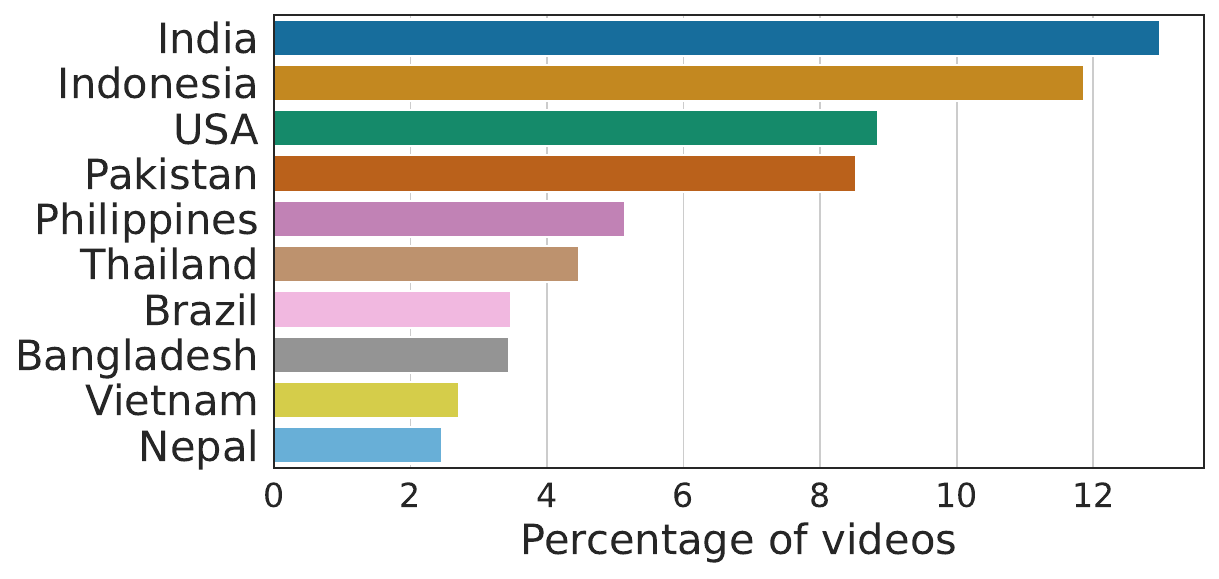}
    \caption{Top 10 regions by prevalence in the dataset with relative percentage of prevalence in the sample.
    India is still the largest one historically, despite the ban in 2020.}
    \label{fig:rq3_barplot}
\end{figure}

\begin{figure}[!t]
    \centering
    \includegraphics[width=1\linewidth]{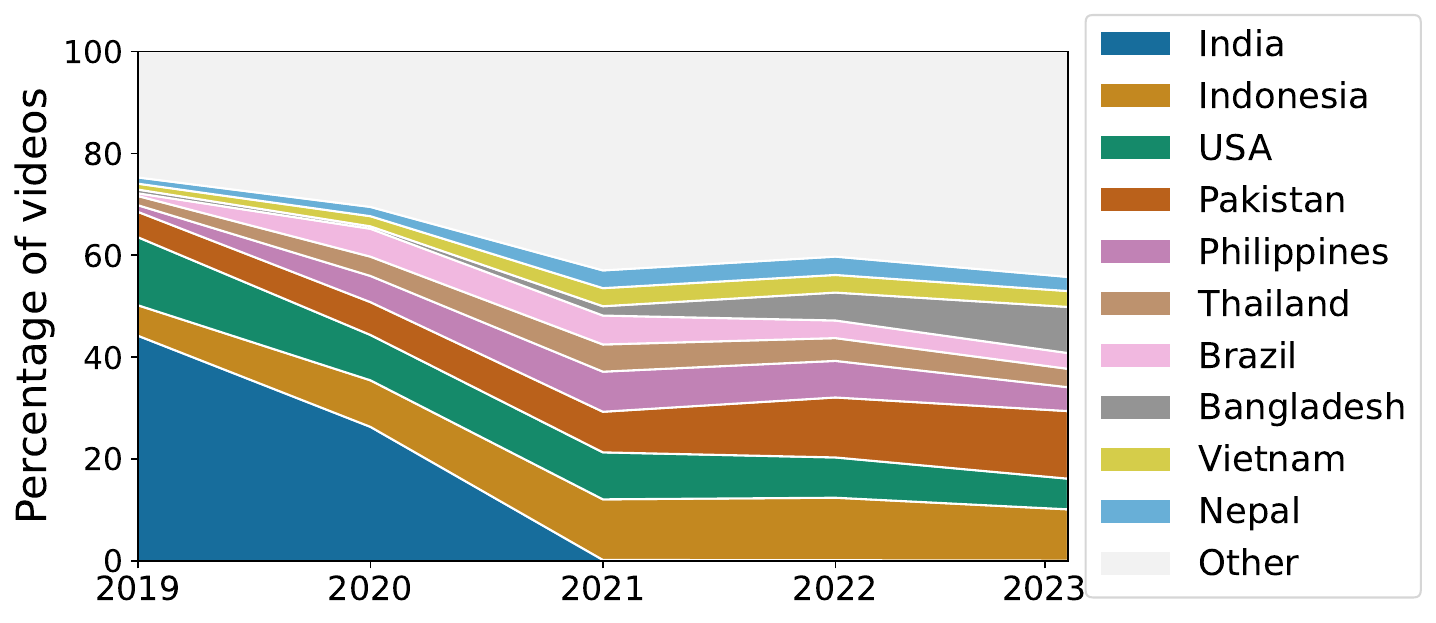}
    \caption{Yearly prevalence of the top 10 regions in our sample. The light-grey area represents all the other regions collected. Most countries in the top 10 are in Asia.}
    \label{fig:rq3_stackplot}
\end{figure}

\Cref{fig:rq3_barplot} shows the top 10 countries by number of videos in our sample.
The first is India, with over 12\% of videos, followed by Indonesia and then the US, which is also the only Western country in the top-10 list. 
We further investigate this aspect by plotting the yearly prevalence of the top 10 countries over the span of the dataset~\Cref{fig:rq3_stackplot}. 

The most evident feature is the prevalence of videos from India and Southeast Asia in general.
From 2019 to mid-2020, India was the most prominent country in our dataset, with over 40\% of the total videos sampled from early 2019. 
The rapidly descending trend is due to a nationwide censorship policy applied in June 2020 which affected TikTok and other Chinese applications~\cite{song2023can}.
Note how the top 10 countries represent just over 60\% of the total videos sampled from the API, thus indicating again a heavy-tailed distribution.

\subsection{Effects of viral hashtags}

\begin{table}[!t]
\caption{Top 10 hashtags by frequency of use in our sample and their virality (manual assessment).}
\label{tab:rq4_hash}
\hfill{}
\begin{tabular}{ll}
\toprule
Hashtag    & Viral? \\
\midrule
fyp        & Yes    \\
foryou     & Yes    \\
duet       & No     \\
capcut     & No     \\
foryoupage & Yes    \\
\bottomrule
\end{tabular}
\hfill{}
\begin{tabular}{ll}
\toprule
Hashtag    & Viral? \\
\midrule
fyp\begin{CJK}{UTF8}{min}シ\end{CJK}      & Yes    \\
viral      & Yes    \\
tiktok     & No     \\
parati     & Yes    \\
trending   & Yes    \\
\bottomrule
\end{tabular}
\hspace{0.1\columnwidth}
\end{table}

\begin{figure}[!t]
    \centering
    \includegraphics[width=1\linewidth]{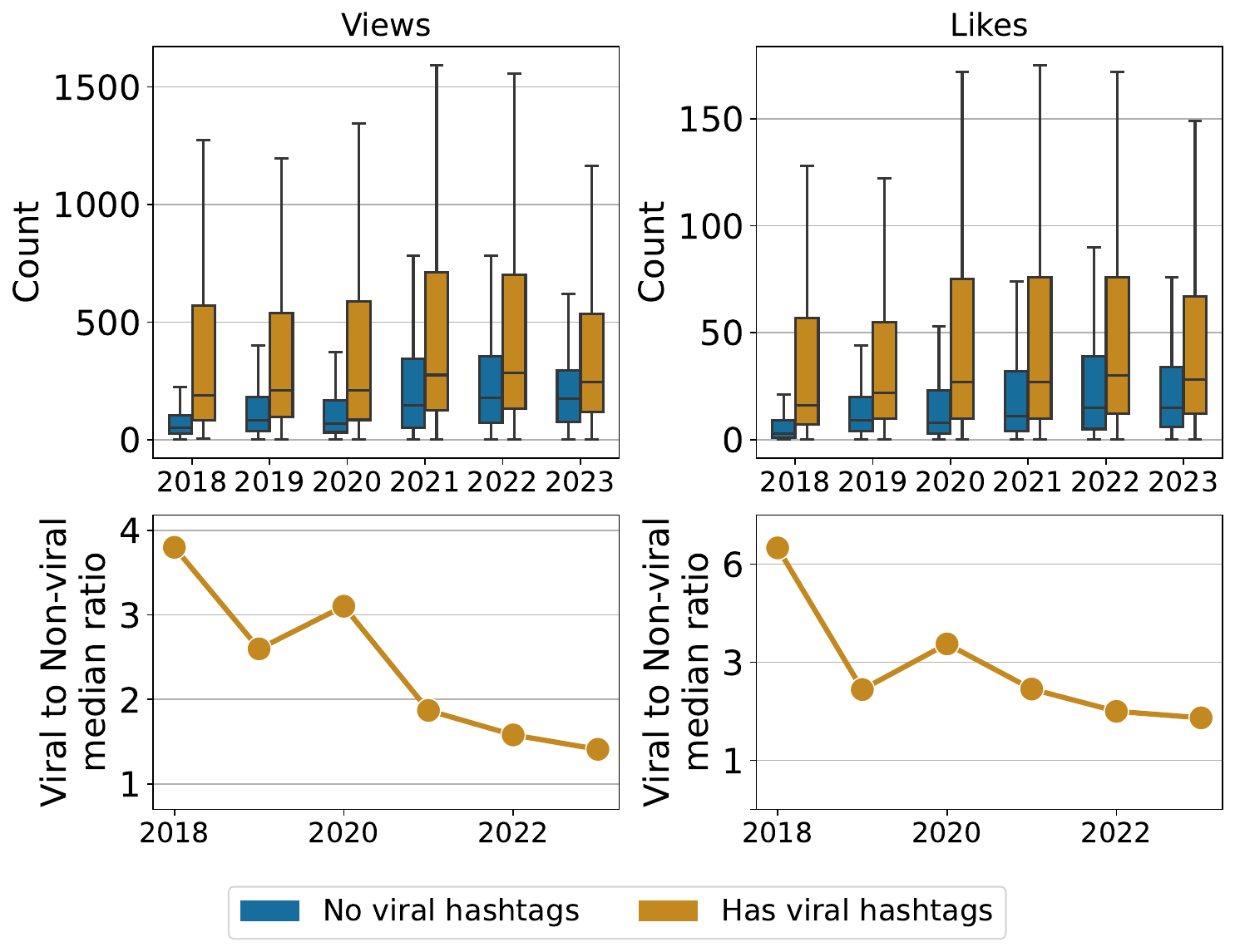}
    \caption{(Top) Yearly distributions of views and likes of videos according to whether they use `viral' hashtags.
    (Bottom) Yearly ratio of the medians of views and likes for videos that use `viral' hashtags vs. those that do not.}
    \label{fig:rq4_boxmedian}
\end{figure}

\begin{figure}[!t]
    \centering
    \includegraphics[width=1\linewidth]{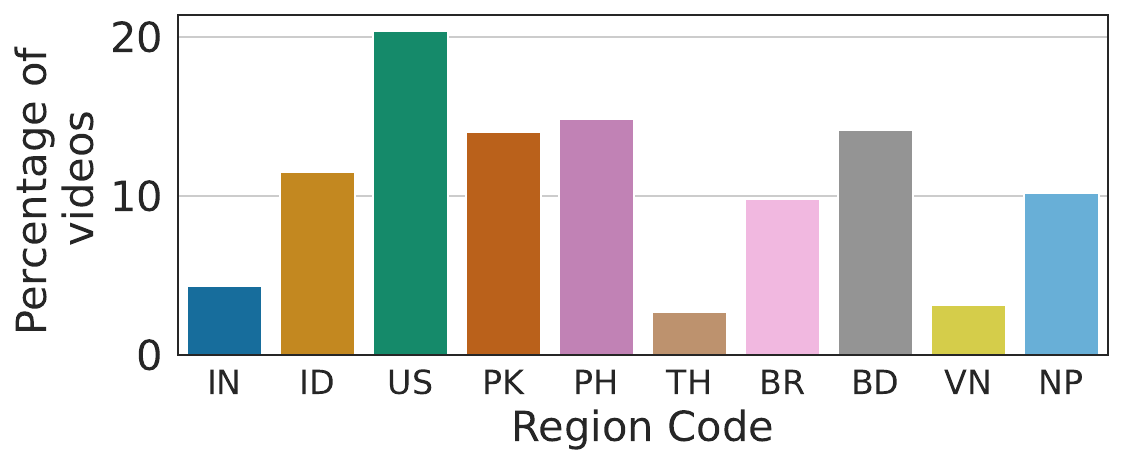}
    \caption{Percentage of videos that use viral hashtags in the top 10 countries by prevalence.}
    \label{fig:rq4_barplot}
\end{figure}

Since their creation on Twitter in 2007, hashtags have morphed into fundamental and pervasive elements of social media culture \cite{bernard2019theory}.
Originally designed for categorization and conversation facilitation, hashtags now play a fundamental role in content discovery and trendsetting across various platforms.
TikTok is no different in this aspect.
Users make use of hashtags to define and aid the discovery of the content they post.
Some of these hashtags are especially employed because they allegedly boost the visibility of the content, by exposing it to the TikTok recommendation algorithms.

\Cref{tab:rq4_hash} shows the top 10 most frequent hashtags used in our sample, with a manual classification of the intent of virality of the hashtag.
This classification is based on the perceived purpose of the hashtags to recall a particular functionality of the social platform: the `For You Page', where the average user of TikTok spends over 60\% of their time~\cite{StokelWalker2020TikToksGS}.
The use of this hashtag indicates the will of the author of the video to `invite' the algorithm to show their content on the `For You Page', thus potentially widening their audience.
\Cref{fig:rq4_boxmedian}~(top) tests this effect by comparing the distribution of views and likes for videos that use at least one `viral' hashtag in their description (approximately 15\% of the total) to the rest of the videos which do not use these hashtags. 
Videos that use `viral' hashtags have significantly more views and likes compared to the ones that do not use them (Two-sided Mann-Whitney, $\mathrm{p}<0.001$).
This behavior is present throughout all the years considered in our study, but if we observe the ratio between the medians of the two distributions (\Cref{fig:rq4_boxmedian}, bottom) we see that the trend tends to decrease in the more recent years.
This result suggests a possible adjustment of the recommendation algorithm to give less weight to the presence of these hashtags.
\Cref{fig:rq4_barplot} shows the top 10 regions by prevalence with the relative percentage of videos that use `viral' hashtags.
It is noteworthy that even non-English speaking countries make use of English hashtags.
This result possibly indicates the intent of the authors to reach an international audience.
However, researchers should take care of potential biases when searching the API with specific English hashtags.

\subsection{Prevalence of Conspiracy Theories}
\begin{figure}
    \centering   \includegraphics[width=1\linewidth]{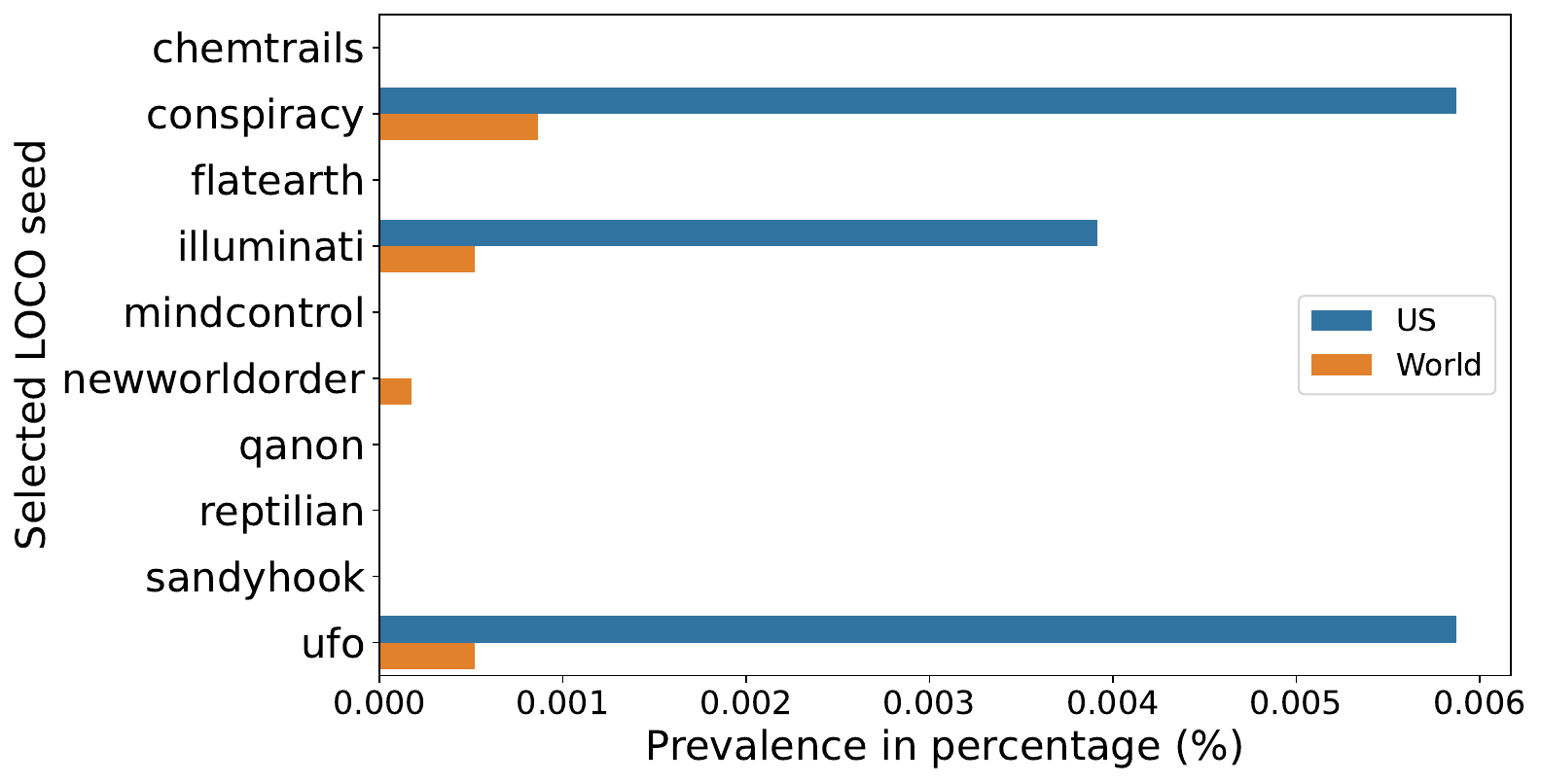}
    \caption{Percentage of videos in the world and in the US that use conspiracy-related hashtags.}
    \label{fig:rq5_barplot}
\end{figure}

We are also interested in giving a preliminary outlook on the presence of hashtags that are related to mainstream conspiracy theories. 
We employed the LOCO dataset by \citet{miani2021loco} to obtain a list of the most prevalent conspiracy seeds (keywords), focusing on the top 20 seeds ( which describe approximately 45\% of the articles on LOCO). We then filtered manually this list, to remove those keywords that were too generic (e.g. \textit{5g}, \textit{coronavirus}, \textit{climatechange}, \textit{barackobama} ...) and keep only the ones that are widely recognized as conspiracy theories.
This resulted in a list of nine seeds of conspiracy theories that we employed as hashtags to search into our dataset plus the keyword `conspiracy' as an additional check.
We show the results of this search in \Cref{fig:rq5_barplot}.

Only three of the nine seeds are present in the dataset, with very low percentages, with the hashtag `conspiracy' being slightly more prevalent.
The prevalence is higher if we focus on the `US' region, since the hashtags themselves are in the English language.
We also compute the \num{99}\% confidence intervals of these percentages via Clopper-Pearson method, which resulted in the order of $10^{-5}$, thus not visible in the plot.
Considering that the dataset we collected gathers videos from all around the world and that the number of videos posted on TikTok amounts to tens of billions ,\footnote{\url{https://www.tiktok.com/transparency/en-us/community-guidelines-enforcement-2023-4/} accessed on 10/03/2024} this analysis estimates that the number of conspiracy videos could amount to hundreds of thousands on a worldwide scale.
\section{Conclusions}

We collected a monthly-based stratified random sample of videos posted on TikTok by using the official Research API.
We then provided a series of relevant insights and statistics about the performance of the API service and the data we obtained, especially oriented to the researchers who are planning to use this service.
Our results describe a significant inability of the API to meet the quotas of requested videos, with a possible internal problem of data quality since querying videos from 2018 provided much fewer results compared to other periods.
We then showed the growth of likes, views, comments, and shares over time, while also providing informative statistics about the global demography of the social media platform.
Researchers should particularly pay attention to the latter, since the majority of videos on the platform originate from Asian countries, and authors in those countries also employ English-language hashtags.
Finally, we showed that the videos that use typical engagement-oriented `viral' hashtags have statistically more views and likes compared to the rest of the sample.

As with any empirical work, our research is subject to limitations.
First of all, our sample is not uniformly random through the 6 years we set for the collection since it is stratified by month. The probability of a video being sampled from all the videos posted on TikTok in six years is much lower than the probability of the same video being sampled in the month of its creation. 
Moreover, the TikTok Research API service is a black-box system and we cannot explore the inner mechanisms that provide us with these supposedly-random results.
This issue compounds with the lack of transparency for what concerns the removed content on this platform, which is currently inaccessible to researchers.

The presence of an official research API by TikTok has opened several research possibilities.
It is now possible to study discourse quality on TikTok and examine several problems that afflict other platforms such as disinformation and coordinated inauthentic behavior.
For instance, looking at their prevalence in these samples can give us an estimate of their presence on the whole social media.
Still, the service offered is far from being ideal, due to the limited number of available requests and the convoluted documentation.

%% The next two lines define the bibliography style to be used, and
%% the bibliography file.
\bibliographystyle{ACM-Reference-Format}
\bibliography{arxiv}

%% If your work has an appendix, this is the place to put it.
\end{document}